\documentclass[journal=jacsat,manuscript=article]{achemso}

\usepackage{chemformula} 
\usepackage[T1]{fontenc} 

\usepackage{graphics}
\usepackage{graphicx}
\usepackage{dcolumn}
\usepackage{bm}
\usepackage{xcolor} 
\usepackage{hyperref}
\author{Z. S. Momtaz}
\email{zahra.sadre-momtaz@neel.cnrs.fr}
\affiliation{NEST, Instituto Nanoscienze CNR and Scuola Normale Superiore, Piazza S. Silvestro 12, I-56127 Pisa, Italy}
\altaffiliation{Current address: Institut N\'eel, CNRS, 25 Av. des Martyrs, 38000 Grenoble, France}
\author{S. Servino}
\affiliation{Department of Physics “E.Fermi”, Universit\`a di Pisa, Largo Pontecorvo 3, I-56127 Pisa, Italy}
\author{V. Demontis}
\author{V. Zannier}
\author{D. Ercolani}
\affiliation{NEST, Instituto Nanoscienze CNR and Scuola Normale Superiore, Piazza S. Silvestro 12, I-56127 Pisa, Italy}
\author{F. Rossi}
\affiliation{IMEM-CNR Institute, Parco Area delle Scienze, Parma, Italy}
\author{F. Rossella}
\author{L. Sorba}
\author{F. Beltram}
\author{S. Roddaro}
\affiliation{NEST, Instituto Nanoscienze CNR and Scuola Normale Superiore, Piazza S. Silvestro 12, I-56127 Pisa, Italy}
\alsoaffiliation[Universit\`a di Pisa]
{Department of Physics “E.Fermi”, Universit\`a di Pisa, Largo Pontecorvo 3, I-56127 Pisa, Italy}
\email{stefano.roddaro@unipi.it}

\title[An \textsf{achemso} demo]
  {Orbital Tuning of Tunnel Coupling in InAs/InP Nanowire Quantum Dots}

\abbreviations{IR,NMR,UV}

\begin{document}

\begin{abstract}
  We report results on the control of barrier transparency in InAs/InP nanowire quantum dots via the electrostatic control of the device electron states. Recent works demonstrated that barrier transparency in this class of devices displays a general trend just depending on the total orbital energy of the trapped electrons. We show that a qualitatively different regime is observed at relatively low filling numbers, where tunneling rates are rather controlled by the axial configuration of the electron orbital. Transmission rates versus filling are further modified by acting on the radial configuration of the orbitals by means of electrostatic gating, and the barrier transparency for the various orbitals is found to evolve as expected from numerical simulations. {\color{black} The possibility to exploit this mechanism to achieve a controlled continuous tuning of the tunneling rate of an individual Coulomb blockade resonance is discussed}.
\end{abstract}

Heterostructured InAs/InP nanowires (NWs) represent an ideal platform for the implementation of single-electron transistors~\cite{bjork2002, bjork2004, fuhrer2007, roddaro2011, romeo2012, nilsson2018,nilsson2017} and a variety of quantum devices for single-photon emission~\cite{dalacu2012}, spin manipulation~\cite{ bjork2005, romeo2012, rossella2014}, thermoelectric conversion~\cite{matthews2012, josefsson2018, prete2019} and more~\cite{nylund2016}. These bottom-up nanostructures can indeed be grown with a fine control on the chemical composition along the NW axis and with atomically-sharp interfaces~\cite{bjork2002, zannier2019, zannier2017, ercolani2009}; this, in combination with the small band mass and favorable Fermi pinning in InAs-based NWs, makes it possible to tightly confine electrons and yields nanostructures characterized by large confinement and charging energies~\cite{bjork2004, romeo2012}. This growth control can be exploited to carefully tailor tunnel barriers properties {\color{black} but leads to tunnel couplings that are less obvious to tune with respect to the ones obtained in gated structures}. This limitation hampers the successful exploitation of these nanostructures in many device architectures: tunnel coupling must be matched to the thermal energy scale for optimal thermoelectric conversion~\cite{nakpathomkun2010,josefsson2018, prete2019}; single-photon detectors based on quantum dots (QDs) can require a wide range of different tunneling rates~\cite{kleinschmidt2006}; the investigation of quantum and manybody phenomena such as those involving the Kondo effects~\cite{goldhaber1998, cronenwett1998, schmid2000, sasaki2000,van2000}, electron spectroscopy for the investigation of proximity effects~\cite{junger2019,doh2005,deng2016} typically require the fine tuning of high-quality tunnel barriers. 

\begin{figure}[ht!]
\begin{center}
\includegraphics[width=0.49\textwidth]{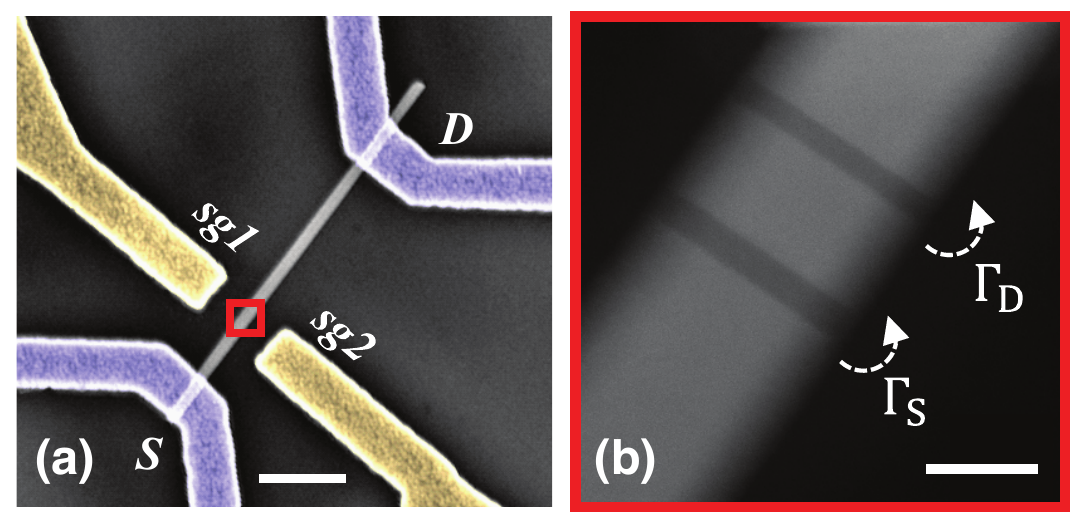}
\caption{{\bf Device architecture.} (a) The studied devices are fabricated on top of doped SiO$_2$/Si and feature source and drain Ohmic contacts (blue) and two lateral field-effect gates (yellow). (b) Z-contrast STEM image of the core of the device, which is a heterostructured InAs/InP nanowire embedding two InP barriers defining a $19\pm 1\,{\rm nm}$-thick InAs island. Scale bars in the two panels correspond to $400\,{\rm nm}$ and $20\,{\rm nm}$, respectively. }
\end{center}
\end{figure}

\begin{figure*}[ht!]
\begin{center}
\includegraphics[width=0.99\textwidth]{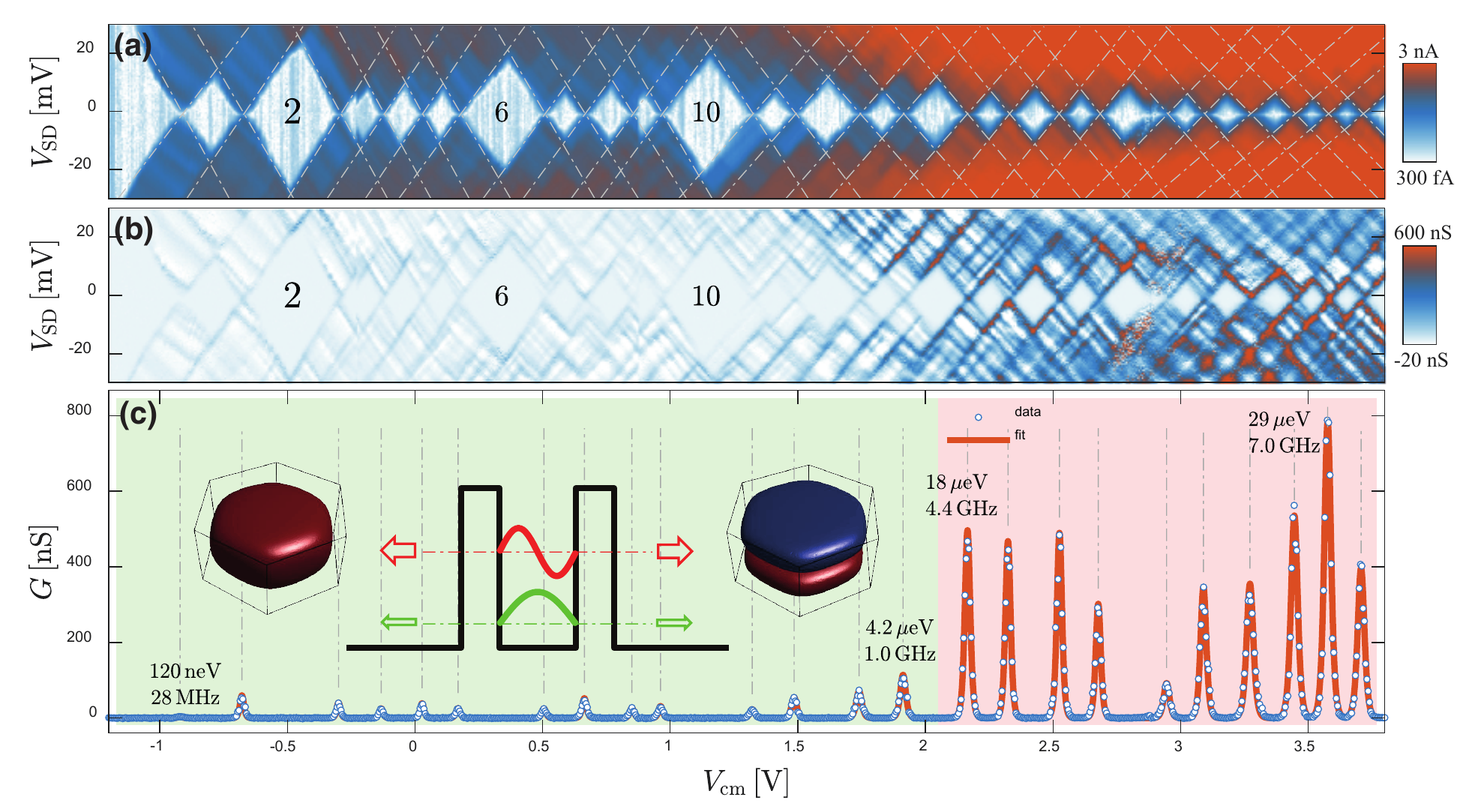}
\caption{\label{fig:CBdiamonds} {\bf Orbital dependence of the tunnel coupling.} (a) Coulomb blockade diagram as a function of the common-mode side gate voltage $V_{\rm cm}$ for one of the studied devices. The color-plot displays the absolute current in logarithmic scale and covers a fairly wide range of fillings going from $N=0$ up to beyond $20$ electrons. (b) The equivalent plot of the differential conductance $dI/dV$ in linear scale highlights the presence of a conductance threshold for $V_{\rm cm}\approx 2\,{\rm V}$ at $N=14$. (c) The zero-bias differential conductance $G$ is fitted using a standard single-level lineshapes, yielding a good agreement with the data with a temperature $T=4.14\,{\rm K}$. As discussed in the main text, the presence of a conductance threshold can be understood in terms of the population of orbitals with different axial quantum numbers, as schematized in the overlay.}
\end{center}
\end{figure*}

A strategy to achieve barrier-transparency tunability in these nanostructures can be based on the control of the orbital mediating conduction through the QD~\cite{thomas2019,barker2019}. In quantum well (QW) systems, tunnel coupling strongly depends on the subband index of the tunneling electrons, but not on its in-plane dynamics since the transverse momentum is conserved in the transmission process~\cite{luryi1985}. In heterostructured NWs, electronic states are fully confined (both axial and radial directions) and a similar dependence may be expected as long as they can be roughly factored in a radial and axial part: the former would be conserved during tunneling through a clean barrier and only the axial configuration of the orbital would matter. This is at odds with what is experimentally observed and tunnel coupling was recently reported to display an overall trend consistent with a relatively simple monotonic dependence on the total orbital energy~\cite{thomas2019}. Here, we show that tunnel rates in InAs/InP QDs can display a qualitatively different behavior in the low filling regime and that this is consistent with  radial orbital-configuration conservation during tunneling. In our experiments, we use devices embedding a single InAs/InP QD and show that tunneling rates typically display a stepwise increase above a given filling threshold. We show that this is caused by the occupation of orbitals belonging to higher-index axial subbands that are naturally characterized by a larger barrier penetration. Our conclusions are crucially supported by the experimental configuration we adopted that allows us to separately control the radial confinement in the QD. This in turn allows us to alter the sequence of weakly/strongly coupled orbital states. {\color{black} As argued in the data discussion, our results indicate that a {\em continuous} tuning of the tunneling rate of individual Coulomb blockade peaks could be in principle achieved and controlled by field effect}. 

The structure of the devices used in the experiment is visible in Figure~1a. Fabrication started with the growth of heterostructured InAs/InP NWs by chemical beam epitaxy (CBE), using Au nanoparticles obtained by thermal dewetting of a thin Au film on InAs (111). The nanostructures have a nominal corner-to-corner ``diameter'' of $48\pm\,5{\rm nm}$ and embed two $5\pm1\,{\rm nm}$ InP barriers separated by a $19\pm 1\,{\rm nm}$ InAs island; {\color{black} see the scanning transmission electron micrograph in Figure~1b obtained using a high-angle annular dark field  (HAADF) detector and depicting a NW nominally identical to the ones used to fabricate the devices}. The broad scattering in the NW parameters is due to the size distribution of the Au nanoparticles obtained by thermal dewetting; significantly sharper distributions are obtained using other methods~\cite{messing2010, gomes2015, zannier2019}. In the scanning electron micrograph, the NW is deposited on top of a degenerately-doped Si substrate covered by $300\,{\rm nm}$ of SiO$_2$. Device fabrication is completed by thermal evaporation of two Ti/Au ($10/100\,{\rm nm}$) ohmic contacts acting as the source (S) and drain (D) electrodes (blue in Figure~1a). In addition, two $200\,{\rm nm}$-wide local side-gate electrodes named sg1 and sg2 (yellow) are aligned at the two sides of the NW, in correspondence to the InAs/InP heterostructure. Further details about device structure and measurement set-up are reported in the Supplementary Information.

Figure~2 reports experimental data and shows that the evolution of tunnel coupling is not trivially connected to the total orbital energy. The two lateral gates and the Si substrate can be used to control the number of electrons ($N$) in the QD and the local electrostatic environment. In particular, the substrate gate is used to enhance carrier density in the NW and to ensure that the source and drain sections of the device display a robust conductance; in the experiments reported in the paper the substrate is held at a potential $+5.5\,{\rm V}$. The lateral electrodes can control the value of $N$ and the QD spectrum, depending on the specific bias configuration. In Figure~2a, sides gates are biased at the same voltage $V_{\rm cm}$ -- i.e. they are operated in a {\em``common mode''} -- and control the QD filling. The colorplot reports the absolute current value in logarithmic scale as a function of $V_{\rm cm}$, starting from pinch-off ($N=0$) up to a filling exceeding $20$ electrons. A typical even-odd filling pattern is obtained, as expected from spin-degeneracy in the quantum Coulomb-blockade regime. Charging energies and level spacing can be extracted from the height of odd and even diamonds. Further data analysis is reported in the Supplementary Information and yields an average charging energy $E_C\approx 5\,{\rm meV}$, level spacings in the range $\Delta \varepsilon=0-15\,{\rm meV}$ and the common-mode lever arm $\alpha_{\rm cm} = 0.04-0.08\,{\rm meV/V}$. The observed energy scales are compatible with prior experimental reports~\cite{bjork2004, romeo2012, johnson2005, salfi2010} with the exception of the side-gate lever arm that strongly depends on the specific electrode geometry used in this experiment. Differential conductance data $dI/dV$ in Figure~2b highlight the presence of a clear threshold in the QD conductance when $N > 14$.

Figure~2c displays the corresponding zero-bias differential conductance $G=dI/dV|_0$: we fit the data using the lineshape expected for a non-degenerate delta-like resonance and estimate the total serial tunnel rate $\Gamma = \Gamma_S\Gamma_D/(\Gamma_S+\Gamma_D)$, where $\Gamma_{S/D}$ are the source and drain barrier couplings~\cite{ihn2010}. {\color{black} The tunnel amplitude is in this case simply linked to the peak amplitude, while the broadening is dominated by thermal effects.} The studied devices display Coulomb peaks with $\Gamma$ values that stochastically change from orbital to orbital, but fall in the range of few hundred neV (green). From $V_{\rm cm}\approx 2.0\,{\rm V}$, a transition to more strongly-coupled resonances is observed (in the range of tens of $\mu{\rm eV}$, red), with an abruptness depending on the specific device tested. The observed step-wise pattern of the tunneling rates is obviously inconsistent with a mere stochastic dependence of the tunneling on the individual orbitals, moreover it is also not simply linked to the total electron energy, as previously observed for QDs with similar nominal dimensions operated at much larger electron filling values~\cite{thomas2019}. In fact here we focus on a complementary regime close to the QD pinch-off: as we shall argue in the following, our data indicate a different phenomenology for which the sudden increase in the tunnel coupling is driven by the occupation of orbitals belonging to the second axial subband in the region confined between the two InP barriers, as sketched in the overlay to Figure~2c. This interpretation is supported by an accurate experimental analysis of the QD filling and energy spectrum, also as a function of an adjustable radial confinement.

\begin{figure}[t!]
\begin{center}
\includegraphics[width=0.49\textwidth]{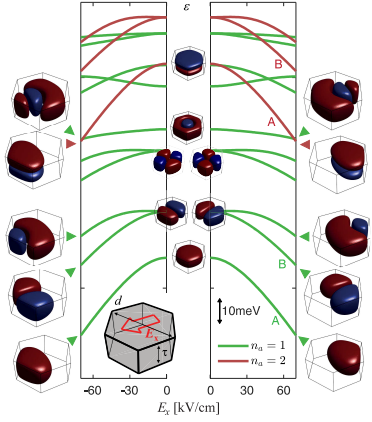}
\caption{\label{fig:epsart} {\bf Orbital states in a hexagonal InAs island.} Single-particle calculation of the orbital states in the InAs/InP QD as a function of an electric field applied along the $x$ direction (see inset sketch). The confinement potential is approximated as a hard-wall hexagonal box with a ``corner-to-corner'' diameter $d=48\,{\rm nm}$ and a thickness of $\tau=19.5\,{\rm nm}$. Lowest-laying orbitals display a decreasing energy at large $E_x$ and are confined to the hexagon corner of lowest potential energy. All the states marked in green correspond to wave-functions originating from the radial confinement of the first axial subband; higher energy copies of the same spectral lines are visible in the red sequence and have two lobes in the axial direction. In this approximation, tunnel rates only depend on $n_a$ and green orbitals are expected to display a smaller tunnel coupling with respect to the red ones.}
\end{center}
\end{figure}

In order to discuss the observed evolution of the electron tunneling rates as a function of $N$, we report in Figure~3 the result of a simulation of the orbital states in the InAs/InP island. Orbitals are calculated using a single-particle approximation and assuming a hexagonal InAs box with an axial thickness $\tau = 19.5\,{\rm nm}$ and a corner-to-corner ``diameter'' $d=48\,{\rm nm}$ (see the sketch inset in the bottom left corner). If the radial confinement potential is large with respect to the energy scales of the problem, one can factor the wavefunction in terms of a radial and axial components labeled by radial and axial quantum numbers $n_r$ and $n_a$ (see Supplementary Information). In particular, {\color{black} assuming a hard-wall potential}, the total energy turns out to be equal to 

\begin{equation}
\varepsilon \approx \varepsilon_{n_r}+\varepsilon_{n_a}=\varepsilon_{n_r}+\frac{h^2n_a^2}{8m^*_e\tau^2}
\end{equation}

\noindent where $n_a$ is a positive integer and $m^*_e=0.04m_e$ is the effective mass in wurzite InAs. In this ideal textbook limit, which turned useful in the interpretation of previous experiments~\cite{rossella2014, rossella2014triple}, the complete QD spectrum is thus expected to contain many shifted copies of the radial excitations spectrum $\varepsilon_{n_r}$, one for each value of the axial quantum number $n_a$. This is clearly visible in the evolution of the spin-degenerate orbital energy $\varepsilon(E_x)$ as a function of an electric field $E_x$ applied in the $x$ direction in the radial plane of the QD (see Figure~3). For instance, the $n_a=1$ and $n_a=2$ states labeled as ``A'' have the same radial configuration and have a similar evolution as a function of $E_x$; an equivalent correspondence is visible for each radial state (the ones labeled as ``B'' and so on). Importantly, as long as the conservation of transverse momentum holds, the tunnel rate depends only on $n_a$: all the $n_a=1$ orbitals are expected to display the same tunnel coupling to the leads, $n_a=2$ ones should exhibit a larger tunneling probability owing to their sizably larger $\varepsilon_{n_a}$ and thus to their larger barrier penetration. {\color{black} We note that the density of states in the NW leads can in principle also play a role, but it is typically found to mostly give rise to mesoscopic fluctuations and is neglected in the current analysis.} This observation provides a first rationalization of the observed device behavior: similar resonance amplitudes are observed as long as only $n_a=1$ orbitals are populated, while the filling of those derived from the $n_a=2$ subband leads to a larger conductivity through the QD. The {\color{black} experimentally} observed threshold $N\approx 10-20$ {\color{black}(see Supplementary Information)} is consistent with the $\tau/d\approx 0.5$ ratio in our QDs. The electric field $E_x$ breaks the symmetry of the radial confinement, lifts level degeneracy and further confines electrons in the radial direction thus modifying $\varepsilon_{n_r}$. This leads to a general enhancement of the energy spacing between radial states in each $\varepsilon_{n_r}$ sequence and thus to crossings between orbitals with different $n_a$. As discussed in the following, this behavior is consistent with our experimental observations and indicates a possible mechanism yielding the controlled smooth tuning of electron tunneling rates in these nanostructures.

In previous works, some of us demonstrated that the energy spectrum of InAs/InP QDs can be strongly modified using a local multiple-gating scheme~\cite{roddaro2011}. To a first approximation, the application of a differential bias between two gates sg1 and sg2 leads to the establishment of an electric field $E_x$ in the QD region. As a result, the energy spectrum is modified. This is experimentally illustrated in Figure~4 where we report the evolution of the measured current as a function of the filling number and of the differential voltage between the two side gates. The results are presented as a color map of the QD current at a bias $1\,{\rm meV}$ as a function of the common-mode voltage $V_{\rm cm}$ (controlling $N$) and of a differential voltage $\Delta V=V_{\rm sg1}-V_{\rm sg2}$ (controlling the transverse electric field $E_x$). The difference $\Delta V$ was distributed on the two side gates according to the equations

\begin{align}
V_{\rm sg1} = V_{\rm cm}+\kappa_1\cdot\Delta V\\
V_{\rm sg2} = V_{\rm cm}-\kappa_2\cdot\Delta V
\end{align}

\noindent where the $\kappa$ parameters satisfy $\kappa_1+\kappa_2=1$. The values of $\kappa_1$ and $\kappa_2$ are chosen so that the average $N$ does not depend on $\Delta V$ and an approximately symmetric evolution is observed for positive and negative values of $\Delta V$. In the specific case shown here, this leads to $\kappa_1 = 0.36$ and $\kappa_2 = 0.64$, the asymmetry being due to a non-intentional non-symmetric arrangement of the two local gates.

\begin{figure}[h!]
\begin{center}
\includegraphics[width=0.49\textwidth]{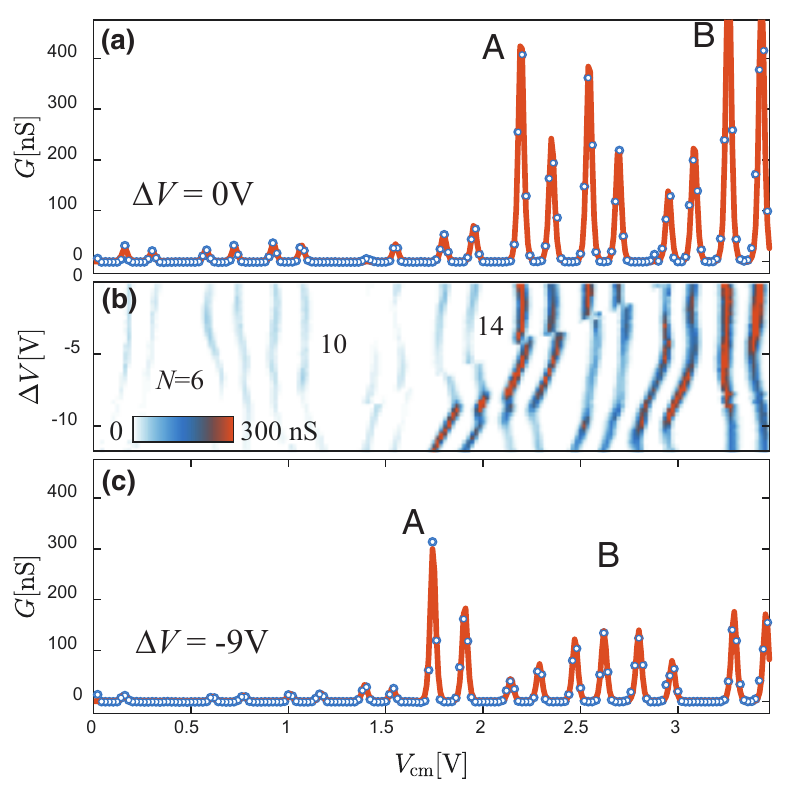}
\caption{\label{fig:epsart} {\bf Spectral modulation and tunnel coupling.} The application of a differential voltage $\Delta V$ to the side gates leads to a significant modulation of the Coulomb peak spacing and of the peak amplitudes. (a) At $\Delta V=0\,{\rm V}$, an onset to stronger Coulomb resonances is observed beyond $N\approx 14$ and large conductivity peaks marked by a letter ``A'' are observed at $V_{\rm cm}\approx 2.2\,{\rm V}$. (b) The application of a finite $\Delta V$ leads to clear crossings between orbitals: in particular, orbital labelled as ``A'' moves to lower energies and the onset to larger conductance peaks shifts to $N=12$. A similar evolution is highlighted by the label ``B''. (c) The cross-section at $\Delta V=-9\,{\rm V}$ clearly shows the new position of the larger conductance peak and the occurence of a strongly non-monotonic evolution of $\Gamma$ as a function of $N$.}
\end{center}
\end{figure}

The initial spectral configuration at $\Delta V=0$ is visible in Figure~4a, where the cross-section as a function of $V_{\rm cm}$ matches the one of the dataset in Figure~2c. As an increasing $\Delta V$ is applied to the side gates, Coulomb blockade peaks display a sizable evolution (Figure~4b) eventually leading to the configuration visible in Figure~4c when $\Delta V=-9 V$ . We note that the threshold to the more strongly coupled orbitals shifts to lower values of $N$, consistently with the evolution in Figure~3. {\color{black} A first rough estimate of the electric field at a given $\Delta V$ can be made based on numerical simulations (see Supplementary Information) and for $|\Delta V|=9\,{\rm V}$ we expect $E_x\approx 45\,{\rm kV/cm}$.} This is in fact a consequence of an energy crossing between QD orbitals, which is clearly visible in the color plot of Figure~4b thanks to the strong orbital dependence of the amplitude of the Coulomb blockade peaks. In particular, the spin doublet A,  visible at $V_{\rm cm}\approx2.2\,{\rm V}$ for zero detuning clearly drifts to lower values of $V_{\rm cm}$, until it crosses another orbital at $\Delta V\approx -9\,{\rm V}$ and is finally located at $V_{\rm cm}\approx 1.8\,{\rm V}$ at the maximum explored value of $\Delta V$. A further strongly coupled peak double B starts at $V_{\rm cm}\approx3.3\,{\rm V}$ and shifts to about $2.5\,{\rm V}$. A similar evolution is observed for positive values of $\Delta V$, as discussed later on.

\begin{figure}[t!]
\begin{center}
\includegraphics[width=0.49\textwidth]{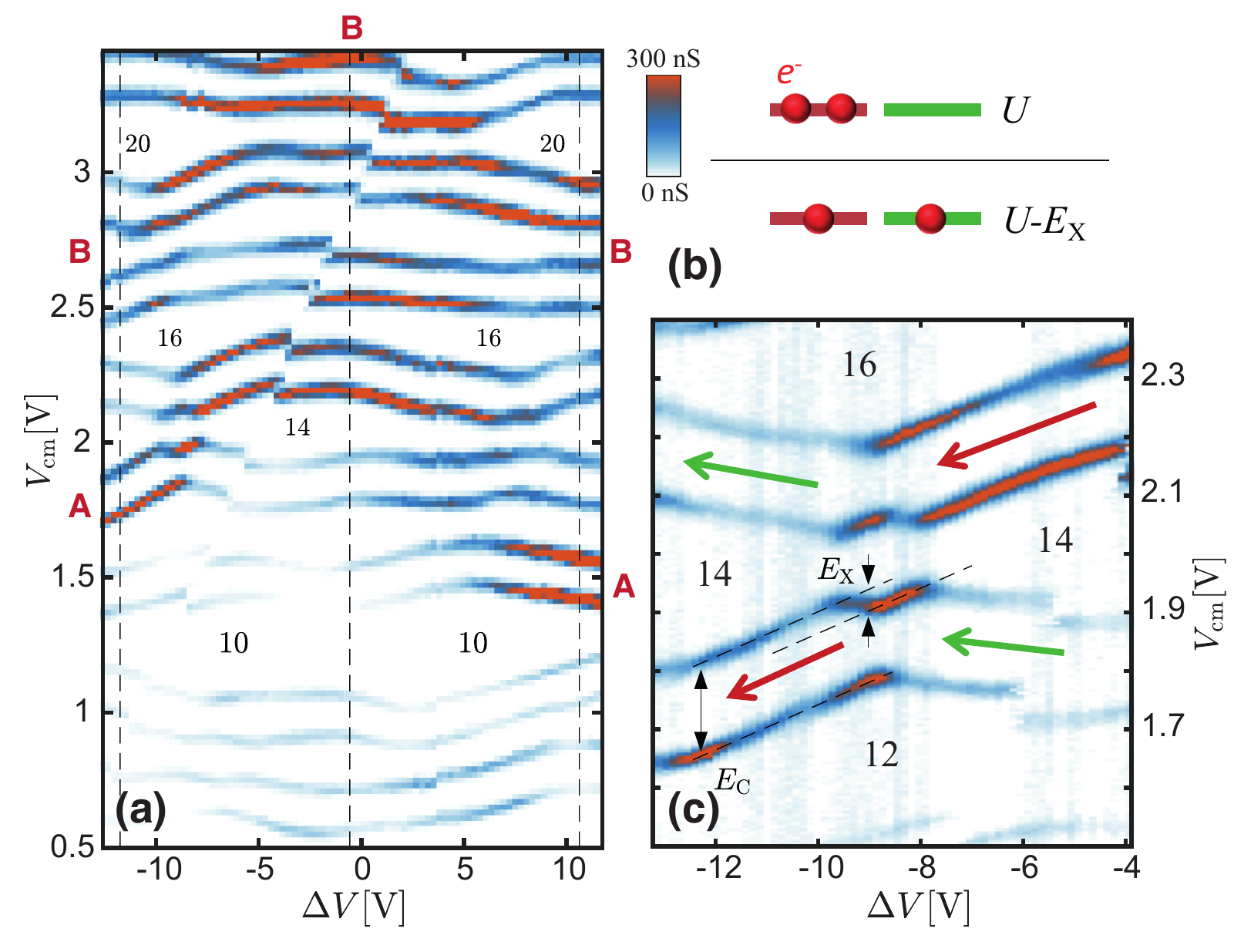}
\caption{\label{fig:epsart} {\bf Overall spectral evolution of the QD.} (a) Full evolution of the Coulomb blockade peaks versus the gate imbalance $\Delta V$ and the common mode voltage $V_{\rm cm}$. Measurements were obtained using a bias of $1\,{\rm mV}$ and cover fillings from $N=6$ to $N=22$. Two resonances with a larger tunnel amplitude are highlighted by the letters ``A'' and ``B'', consistently with Figure~3. (b) Denser current map in correspondence one of the level crossings breaking the even-odd filling scheme. The phenomenology is known to derive from the reduction of the potential energy $U$ of the last two electrons in the QD thanks to a partial filling of two nearly degenerate levels (see sketch on top of the panel), for instance as a consequence of exchange interaction. An energy gain $E_{\rm X}=1.85\pm0.13\,{\rm meV}$ can be determined from the distance in $V_{\rm cm}$ between the dashed lines and from the QD lever arm at $N\approx 14$.}
\end{center}
\end{figure}

The analysis of Figure~4 highlights a few important experimental facts. First of all, the increase in tunneling amplitude $\Gamma$ is not simply monotonic in the orbital energy and thus in $N$. This emerges clearly by looking at the evolution of the Coulomb blockade as a function of $\Delta V$: beyond a threshold, orbitals with different $n_a$ can easily occur at energies that are too close to be resolved and/or are partially mixed, making it difficult to identify the nature of the QD orbitals involved. Differently, considering the overall spectral evolution of Figure~4b, specific orbitals with a stronger tunneling emerge and maintain their (larger) tunnel amplitude throughout the evolution versus $\Delta V$. Peaks with larger tunneling rates tend to shift towards {\em lower} values of $N$ for increasing values of $\Delta V$ (i.e. increasing transverse electric field $E_x$). Such a behavior is consistent with the predictions of Figure~3 and can be understood as a consequence of the radial confinement of the states. We further note that the {\color{black} presence of level crossings between strongly- and weakly-coupled orbitals could -- in the presence of a sufficiently large anticrossing, and thus hybridization, between the two wavefunctions -- lead to a continuous tuning of the Coulomb blockade amplitude versus $\Delta V$.} In the current experiment, the two crossing levels are too close to be resolved so that a suitable {\color{black}anticrossing and hybridization} should be introduced in the system.

The overall evolution of the QD spectrum in the filling range from $N=6$ to $N=22$ is illustrated in Figure~5a{\color{black}, extending the dataset visible in Figure~4b}. The good stability of the studied nano-heterostructures allows the investigation of a fairly large range of gate configurations, but charge rearrangements could not be completely avoided. Beyond the diagonal one visible in the colorplot, a vertical rearrangement at $\Delta V\approx 4\,{\rm V}$ was numerically removed by shifting the value of $V_{\rm cm}$ to improve readability (raw data are reported in the Supplementary Information). Orbitals with larger tunneling are visible in red and clearly reproduce the crossing scheme reported in Figure~3, with an approximate mirror symmetry around the central dashed line. While at $\Delta V\approx 0$ the transition to larger tunneling resonances occurs after $N=14$, at the configurations highlighted by the outer dashed lines the transition shifts to $N=10-12$. We stress that even if only two QD orbitals in Figure~5a exhibit a larger tunneling amplitude, for generic $\Delta V$ values many more CB peaks display an amplified amplitude. This behavior is {\color{black} likely connected to a hybridization between strongly- and weakly-coupled orbitals and leads to intermediate (and not small) tunnel amplitudes. This} highlights that a multigate architecture is needed to study the effect {\color{black} and identify the role of the involved orbitals in the QD}. For selected values of $\Delta V$ (e.g. along the left dashed line) low-tunneling orbitals can be energetically decoupled and CB with a small amplitude can be observed even at relatively large $N$ values, confirming our interpretation. We finally note that a precise matching between experimental data and Figure~3 can not be expected owing to the approximations used: even restricting the analysis to the single-particle picture, band bending (inside the QD and at the NW surface) was not taken into account. 

Figure~5a shows that level crossing at times breaks the typical even-odd filling sequence {\color{black} and cannot be described by a standard constant interaction model}: in Figure~5b, we report a finer map of the one at $N=14$ for $\Delta V<0$. This class of crossings are found to occur particularly frequently when they are associated with a degeneracy between orbitals with different $n_a$. The breakdown of the even-odd scheme was previously highlighted in the literature in association with level crossings induced by magnetic fields~\cite{tarucha2000}, local gating~\cite{fuhrer2003} or -- similarly to the case reported here -- radial confinement~\cite{romeo2012,rossella2014triple}. The effect is fundamentally connected to the existence of a reduction of electrostatic cost $U$ in the case of a partial occupation of two nearly-degenerate orbitals, with respect to the more conventional complete filling of the lowest energy one (see sketch on top of Figure~5b). In our case we estimate a reduction $E_{\rm X} = 1.85\pm0.13\,{\rm meV}$ based on the $V_{\rm cm}$ shift between the dashed line and on the lever arm $\alpha_{\rm cm}=0.050\pm0.0035\,{\rm eV/V}$. The origin of the effect is typically related to exchange interaction~\cite{tarucha2000}, but recent works on similar NW systems have convincingly demonstrated that a major role can also be played by the spatial segregation of the two orbitals, leading to the formation of an effective parallel double QD system with a sizable reduction of direct Coulomb repulsion for the partial filling configuration~\cite{nilsson2018,nilsson2017}. The main difference between the two scenarios is the degeneracy of the manybody configuration. In the exchange scenario, it is reduced by the lifting between $S=1$ triplet and $S=0$ singlet states while a full spin degeneracy is expected when the energy gain of the {\color{black} partial filling of the two nearly-degenerate orbitals} has a purely Coulombian origin. While the effect represents an intriguing aspect of the phenomenology of the nanodevices, our current NWs do not provide a definite answer: on one hand, devices do not display sufficiently clear excited state patterns to establish the presence of a singlet-triplet splitting (further evidences available in the Supplementary Information); on the other one, the studied QD structure does not present obvious features leading to the formation of a parallel double dot at the crossing between states with different $n_a$. We note, however, that the orbital with a larger $n_a$ will be naturally more confined in the radial direction with respect to the one with lower $n_a$, possibly explaining --  within the spatial segregation scenario -- the occurrence of the effect.

In conclusion, we have demonstrated that the evolution of tunneling rates in InAs/InP QDs at small filling numbers displays a filling threshold leading to resonances with larger amplitude. The effect is understood in terms of the occupation of orbital states originating from different axial subbands confined between the two InP barriers and characterized by a larger tunneling probability. Our conclusion critically depends on the possibility to modulate of the QD spectrum and re-order the sequence of orbitals with larger and smaller tunneling amplitudes, leading to a clear identification of the role of the different orbitals. Our work indicates an effective route to continuous and controllable tuning of the tunnel coupling in heterostructured NW systems.\\

{\bf Methods.} InAs/InP heterostructured NWs were grown by chemical beam epitaxy seeded by metallic nanoparticles obtained from thermal dewetting of an Au thin film. Growth was performed at $390\pm10\,{\rm C}$ using trimethylindium (TMIn: $0.3\,{\rm Torr}$, cracked at the NW surface), tert-butylarsine (TBA: $1.0\,{\rm Torr}$ cracked at $1000\,{\rm C}$), and tributylphosphine (TBP: $4.0\,{\rm Torr}$, cracked at $1000\,{\rm C}$). NWs have a wurtzite crystal structure, InAs/InP and InP/InAs interfaces were realized without any interruption by switching group-V precursors. The average position of the QD along the NW and from Au nanoparticle is $500\pm20\,{\rm nm}$ which was estimated based on transmission electron microscopy (JEOL JEM 2200 FS operated at 200 kV), leading to a typical alignment error of $\pm 50\,{\rm nm}$ in fabrication. Ohmic contacts were obtained by thermal evaporation of a Ti/Au (10/100 nm) bilayer, after a chemical passivation step using a  (NH$_4$)S$_{\rm x}$ solution~\cite{suyatin2007}. The orbital configurations in the QD system were simulated using a commercial PDE solver (COMSOL Multiphysics). Transport measurements were performed in a Heliox system running at $4.2\,{\rm K}$, using Yokogawa and Stanford Research System voltage sources and a low-noise DL-1211 current preamplifier.

\begin{acknowledgement}

 S.R. and Z.S.M. acknowledge the financial support of the ``QUANTRA'' project funded by the Italian Ministry of Foreign Affairs. S.R. acknowledges A.Ghirri for useful discussions.  L.S. and V.Z.  acknowledge the partial final support by the ``SUPERTOP'' project (QUANTERA ERANET Cofund Action in Quantum Technologies) and by the ``AndQC'' project (FET-OPEN).
 
\end{acknowledgement}

\bibliography{tunablebarriersNotes}

\newpage
\begin{center}
{\Large\bf Supporting Information:
Orbital Tuning of Tunnel Coupling in InAs/InP Nanowire Quantum dots}
\end{center}


\section*{S.1 Parameters of the device reported in the main text}

Three different devices were fully characterized within this study. The first one, whose results are presented in the main text, will be called hereafter D\#1, and the other two devices will be called D\#2 and D\#3. Figure S1, reports the key experimental parameters of each of the measured devices. Experimental data was fitted using a constant interaction model and the resulting parameters are reported as a function of the filling number N, for every Coulomb blockade (CB) diamond. In particular, we report: (a) the addition energy $\Delta$U; (b) the common-mode gate lever arm, 
$\alpha_{cm}$; (c) the common-mode capacitance $C_{cm}$ and the total capacitance $C_{\Sigma}$ of the quantum dot. The parameters were calculated starting from a fit of the slopes of the Coulomb diamond’s edges of the stability diagram. The $\Delta$U values for odd diamonds (blue dots in the graph) correspond to the charging energies ($E_c$), while the energy difference between even (grey dots in the graph) and odd diamonds can be used to estimate the level spacing ($\Delta\varepsilon$). The charging energy in the QD of D\#1 is about 10 meV, while the level spacing reaches the values up to 10-15 meV. The common-mode lever arm used for the energy conversion in Figure 5 of the main paper was obtained by a linear fit of the data reported above (around N=14).\\

\begin{figure}[ht!]
\begin{center}
\includegraphics[width=0.7\linewidth]{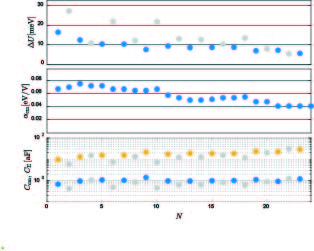}
\caption{Main {\color{black}fit} parameters of device D\#1 (device reported in the main text).}
\end{center}
\end{figure}

The tunneling rate $\Gamma$ for each peak was extracted by fitting the zero bias conductance peaks (Figure  2.c in the main text) using the typical line-shape for a non-degenerate delta-like resonances~\cite {beenakker1991}

\begin{equation}
G = \frac{e^2}{h}\cdot\frac{\hbar\Gamma}{k_{B}T}\cdot\frac{1}{4{\cosh}^2 [\alpha_{k} (V^N_{g}-V_{g})/2k_{B}T]}
\label{eq:1}
\end{equation}
where $\alpha_{k}$ is the gate lever arm,  $V_{g}$ is the gate voltage and $V^N_{g}$ is peak gate voltage at filling N. Corrections can be expected considering the spin degeneracy, but these effects were beyond the scope of the current paper. {\color{black} The results of the described fitting procedure are reported in the following table.}

\begin{table}[h!]
\begin{center}
\color{black}
\begin{tabular}{c|cccccccccccc}
\hline
$N$ &  $1$ &  $2$ &  $3$ &  $4$ &  $5$ &  $6$ &  $7$ &  $8$ & $9$ & $10$ & $11$ & $12$ \\ 
\hline
$\Gamma{\rm (\mu eV)}$ & $0.17$ & $2.2$ & $1.5$ & $0.95$  
                       & $1.4$ & $0.96$ & $0.93$ & $1.9$
											 & $0.98$ & $1.2$ & $0.86$ & $2.0$ \\
$\Gamma{\rm (GHz)}$ & $0.04$ & $0.54$ & $0.36$ & $0.23$ 
                    & $0.35$ & $0.23$ & $0.23$ & $0.47$ 
										& $0.24$ & $0.28$ & $0.21$ & $0.49$\\
\hline
$N$ & $13$ & $14$ & $15$ & $16$ & $17$ & $18$ & $19$ & $20$ & $21$ & $22$ & $23$ & $24$ \\
\hline
$\Gamma{\rm (\mu eV)}$ & $2.6$ & $4.2$ & $18$ & $17$ 
                       & $18$ & $11$ & $3.4$ & $13$ 
                       & $13$ & $20$ & $29$ & $15$ \\
$\Gamma{\rm (GHz)}$ & $0.63$ &  $1.0$ &  $4.4$ &  $4.2$ 
                    & $4.4$ &  $2.7$ & $0.83$ &  $3.1$
                    & $3.2$ & $4.8$ & $7.0$ & $3.5$\\
\end{tabular}
\end{center}
\end{table}

\begin{figure}[h!]
\begin{center}
\includegraphics[width=1\textwidth]{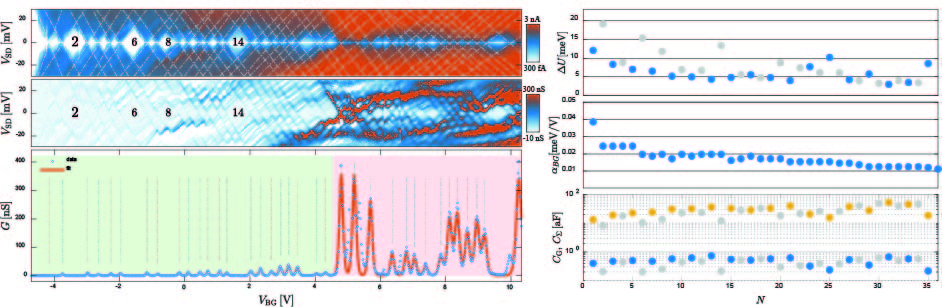}

\includegraphics[width=1\textwidth]{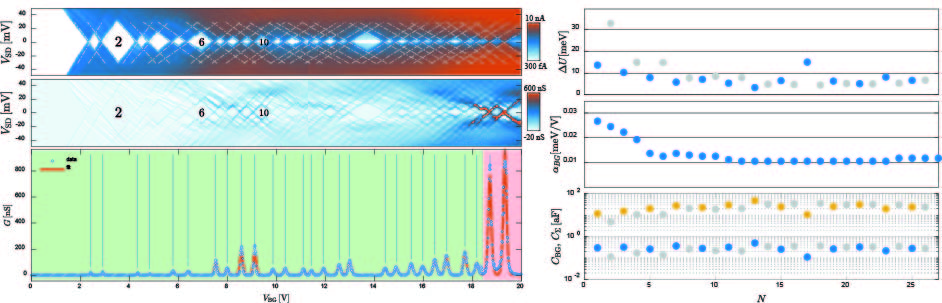}
\caption* {Figure S2. Left: Coulomb blockade diagrams for D\#2 and D\#3 as a function of the back gate voltage, absolute current in log scale and differential conductance at the bottom the zero-bias conductance is reported. Right: parameters extracted from the dataset. Note that the even-odd assignment is not reliable for large filling values due to the presence of charge rearrangements.}
\end{center}
\end{figure}

In Figure S2 we report the stability diagram and extracted parameters for D\#2 {\color{black} and D\#3}. In the plot, the back gate electrode was used to control the electron population inside the QD and an abrupt threshold to larger tunneling was observed. Devices D\#2 and D\#3 displayed a similar tunability using lateral gates but a full investigation of the spectral dependence could not be performed due to the presence of too many charge rearrangements over the large voltage swings necessary to perform the study. {\color{black} The threshold for the observation of large coupling orbitals was observed to be $N=22$ and $N\approx24$ for devices D\#2 and D\#3, respectively. Note the electron filling in D\#3 cannot be reliably identified due to the presence of a non-negligible charge rearrangement.}

\section*{S.2 Numerical models}

\begin{figure}
\begin{center}
\includegraphics[width=0.4\textwidth]{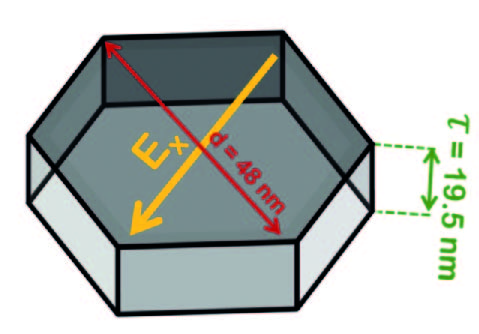}
\caption* {Figure S3: Sketch of the hexagonal box used as a model for the hard-wall NW QD}
\end{center}
\end{figure}

The main target of the simulations was to estimate the expected spacing between radial and axial excitations in the studied QD, i.e. the number of orbitals that need to be filled before larger tunneling resonances can be experimentally observed. To this end, we made a set of simplified assumptions and used a single particle model to predict the expected energy spacings. The nominal geometry of the studied QDs is reported in Figure S3; the InAs island is a hexagonal box with an axial thickness $\tau =$ 19.5 nm in the $z$ direction and a corner-to-corner diameter $d =$ 48 nm.

 Calculations were performed using the PDE solver Comsol Multiphysics and the following further approximations were made: (i) the confinement potential is infinite and barrier penetration is neglected; (ii) band bending at the NW surface and the non-parabolicity effects are neglected. In these approximations, the wave function can be factored as the product $\psi(x, y, z) = A(z)B(x, y)$, where the axial component has a particularly simple expression $A(z) = sin(kz)$ and $k_{z} = \pi n_{a}/\tau$ where $n_{a}$ is an integer. This leads to eigenvalues of the form

\begin{equation}
\varepsilon =\varepsilon_{n_{r}} + \varepsilon_{n_{a}} = \varepsilon_{n_{r}} (E_{x})+\frac{h^2 n_{a}^2}{8 m^*_{e}\tau ^2}
\label{eq:2}
\end{equation}

where $n_{r}$ is the quantum number of the radial problem and the transvers electric field acts only on $\varepsilon_{n_{r}}$ . While the adopted approximations are clearly strong, the resulting estimates on the filling numbers were found to be reliable in the limit of low occupation numbers. Importantly, it has to be noted that in such a limit the transmission across the barrier only depends to the $z$ part of the eigenvalue problem, and thus only on $n_{a}$. In particular, larger transmission amplitudes can be expected for larger $k_{z}$ values and thus for larger $n_{a}$ quantum numbers. This is in full agreement with the experimental observations reported in the main text in Figure 2c.

\begin{figure}
\begin{center}
\includegraphics[width=0.5\textwidth]{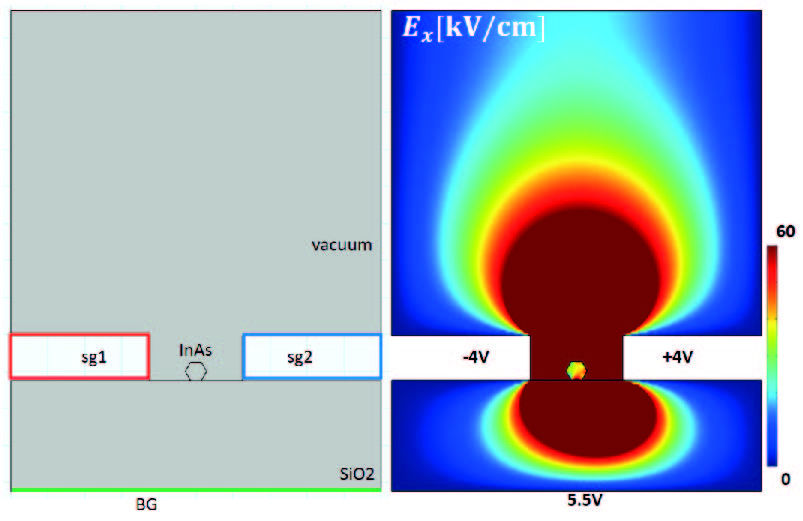}
\caption* {Figure S4: Finite element simulation of the electrostatics of the multi-gate architecture used for this study.}
\end{center}
\end{figure}

In order to allow a semi-quantitative comparison between the single-particle orbital energies reported in the main text with the applied gate voltages, a set of 2D electrostatic simulations have been performed. Figure S4 shows the results for the transverse component of the electric field $E_{x}$ for the following gating configuration: $V_{cm}= 0V$, $\Delta V$ = 8V and $V_{BG}$ = 5.5 V. Known dielectric constants BG for silicon oxide and InAs were used to estimate the local electric field for a gate gap of 250 nm (based on SEM imaging of the measured device, see Figure S7). In the stated configuration, we obtain an average field in the $x$ direction equal to about 40 kV/cm which leads to a consistent comparison between the energy plots reported in Figure 3 and those of Figure 5. {\color{black} The numerical result provides a first rough estimate of the correspondence between lateral gate imbalance  and transverse fields: about 1 kV/cm can be expected for every $\approx 0.2$ V of imbalance between the lateral gates.} We note the calculation does not take into account the (unknown) screening from mobile charges at the surface and in the bulk of the InAs nanowire and thus should be only regarded as a rough consistency check.

\section*{S.3 Raw version of the colormap in Figure 5} 

\begin{figure}
\begin{center}
\includegraphics[width=0.4\textwidth]{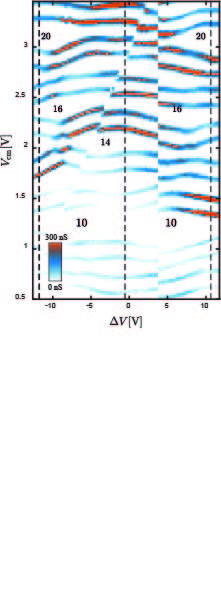}
\caption*{Figure S5: Raw data used for the plot visible in Figure 5 of the main text.}
\end{center}
\end{figure}

Figure 5 in the main text shows the evolution of the Coulomb blockade peaks versus the gate imbalance $\Delta V$ and the common mode voltage $V_{cm}$ . The map required significant voltage swings on the two gates, up to voltages beyond $\pm10V$. Despite the general very good stability of the studied devices, charge rearrangements could not be completely avoided over such large gate swings and imperfections occurred even in the most stable devices. In order to improve the readability of the plot a reproducible charge rearrangement occurring at $\Delta V \approx 4V$ was numerically removed in the color-plot reported in the main text. The original dataset can be seen on the right hand side on Figure S5. Filling number are determined according the correspondence with data in Figure 2.

\section*{S.4 Finite bias data {\color{black} and breakdown of the even-odd filling}}

The physical origin of the {\color{black} breakdown of the even-odd filling sequence} observed in the spectral evolution of Figure 5a, {\color{black} e.g. at the level crossing highlighted} in Figure 5b, could not be conclusively identified in this work. Similar phenomenology has been reported in the literature and has been typically connected the role of exchange interaction in promoting a partial filling of two orbitals and a breakdown of the even-odd filling scheme. Another possibility is that direct Coulomb interaction between the two partially occupied orbitals is simply weaker, because electrons are spatially located in two different regions of the real space. This has been convincingly demonstrated in~\cite{nilsson2018}, where two independent parallel dot are formed in the low filling regime, and it has also been suggested in~\cite{rossella2014}, where the presence of multiple barriers can be expected to lead to a natural localization of the orbitals between different pairs of consecutive barriers. The expected origin of the effect is less obvious to identify in the current device, but one experimental possibility consists in studying the excitation spectrum and look for evidences of a splitting between triplet and single states linked to the different possible spin configurations of the two partially occupied orbitals. The current devices do not provide clear evidence of the presence of such a splitting and no conclusive result could be reached. 

\begin{figure*}[ht!]
\begin{center}
\includegraphics[width=1\textwidth]{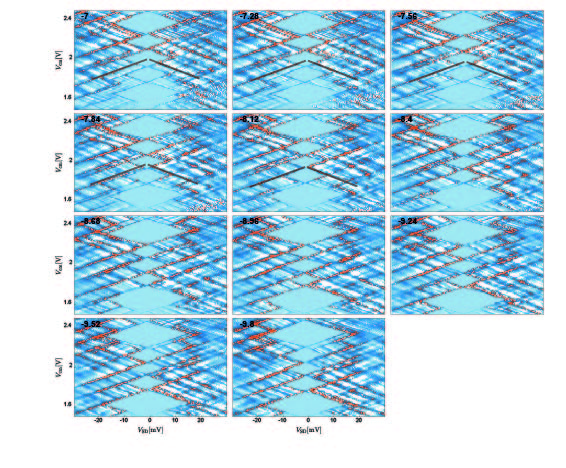}
\caption*{Figure S6: Diamond scans obtained at different cross-sections at fixed detuning $\Delta V$ in the filling range going from N=12 (bottom diamond) up to N=16 (top diamond)}
\end{center}
\end{figure*}

In Figure S6 we report a sequence of diamond scans (differential conductance) as a function of the common-mode gate voltage $V_{cm}$, for a discrete set of imbalance values $\Delta V$ covering the “anomalous” crossing occurring at N=14 for $\Delta V \approx -9 V$. Few excited state lines are visible in measurement but they appear to be connect with the partial or total filling of the available orbitals. For instance, in the sequence starting from imbalance -7V we highlighted an excitation line for the tunneling of the 14$^{th}$ electron. The excitation energy is found to decrease until the size of the N=14 diamond is minimized at imbalance -8.12 V. This voltage marks the beginning of the gating region where the even-odd filling scheme is violated. This excitation line is obviously related to a partial filling of both the crossing orbitals and the process becomes the lowest energy one after imbalance -8.12 V. In the case of exchange interaction, one would expect the presence of both a S=1 (lower energy) and S=0 (higher energy) resonance for such a process. According to the energy gain of 1.86 meV for the partial filling of both orbitals, in the case of exchange the singlet-triplet gap would be expected to be about 3.5 meV~\cite{tarucha2000}. No obvious evidence for such a further resonance could be highlighted.

\section*{S.5 Device structure and measurement set-up}

The structure of the studied devices is shown in Figure S7 along with a sketch of the measurement setup. Different gaps between the gates (GG) have been studied in different devices. The device analyzed in the main text is reported in the SEM picture and had a gate gap of about 250 nm.

\begin{figure*}[ht!]
\begin{center}
\includegraphics[width=0.7\textwidth]{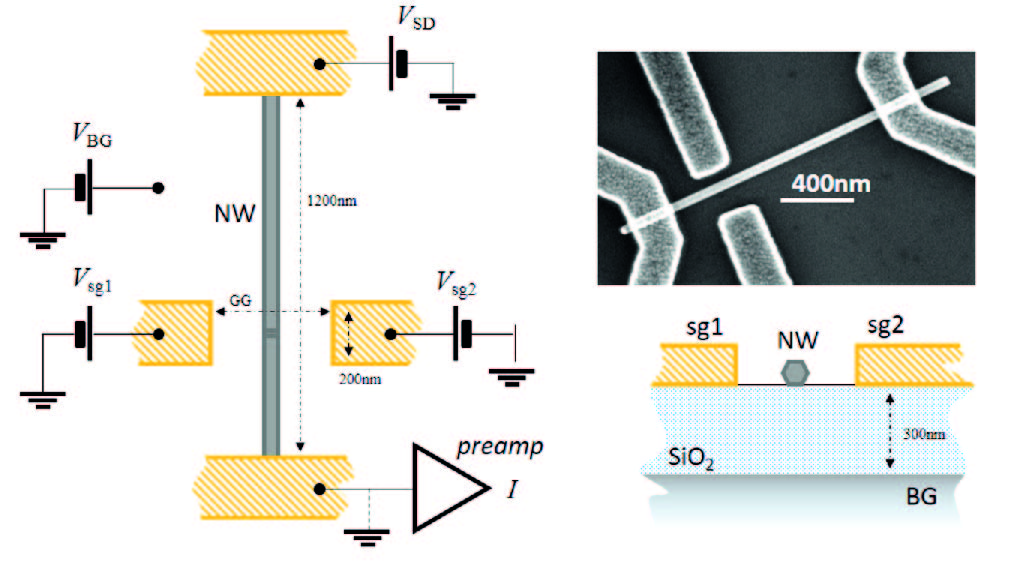}
\caption*{Figure S7: Device structure, SEM picture of the device studied in the main text, and sketch of the measurement setup.}
\end{center}
\end{figure*}

\end{document}